\documentclass[10pt]{iopart}

\usepackage{graphicx}
\usepackage{tabularx}
\usepackage{wrapfig}
\usepackage{booktabs}
\usepackage{siunitx}
\usepackage{parskip}
\usepackage{iopams}
\expandafter\let\csname equation*\endcsname\relax
\expandafter\let\csname endequation*\endcsname\relax
\usepackage{amsmath}
\usepackage{xcolor}
\usepackage{mathptmx}
\usepackage{etoolbox}
\usepackage{comment}
\usepackage{enumitem}
\usepackage[inkscapeformat=png]{svg}
\usepackage{subcaption}
\usepackage{bm}% bold math
\usepackage{optidef} % define optimization problems

\usepackage{hyperref}
\hypersetup{
    colorlinks=true,
    linkcolor=gray,      
    urlcolor=black,
    citecolor=gray
    }
\usepackage[
backend=biber,
style=alphabetic,
sorting=ynt
]{biblatex}
\newcommand{\BeIon}{$^9$Be$^+\,$}
\newcommand{\vecp}[1]{\mkern2mu \vec{#1} \mkern2mu}

\begin{document}

\title{Characterization of Inner Control Electrode Shapes for Multi-Layer Surface-Electrode Ion Traps}

\useSymbolFootnotes
\author{
Florian Ungerechts$^{1}$,
Brigitte Kaune$^{1}$
and Christian Ospelkaus$^{1,2}$}

\address{$^1$ Institut für Quantenoptik,  Leibniz Universität Hannover, Welfengarten 1,30167, Hannover, Germany}
\address{$^2$ Physikalisch-Technische Bundesanstalt, Bundesallee 100, 38116 Braunschweig, Germany}

\ead{ungerechts@iqo.uni-hannover.de}

\vspace{10pt}

\begin{abstract}
Microfabricated surface-electrode traps are a scalable platform for trapped-ion quantum processors. Recent advances in fabrication techniques have enabled the design of increasingly complex multi-layer structures. Yet the control electrodes remain mostly unchanged and of rectangular shape. We systematically analyze asymmetric inner control electrode shapes for simultaneous axial and radial control in multi-layer surface traps, characterize and compare a selection of different shapes, and verify their capabilities in realistic use-case scenarios for ion transport and micromotion compensation. Eliminating the need for the commonly used additional outer control electrodes, asymmetric inner control electrodes increase the compactness and space efficiency of surface-electrode traps while concurrently reducing the number of control signals. The improved control voltage efficiency of using solely inner electrodes enables the device's entire direct-current (DC) supply to be provided by integrated Cryo-CMOS circuits, further enhancing the scalability of the processor.
\end{abstract}

%\submitto{\NJP}
%\maketitle
% 

\ioptwocol

\tableofcontents
\markboth{\leftmark}{\rightmark}

\useArabicFootnotes

\section{Introduction}
\label{sec: Introduction}
Trapped ions are a leading implementation for quantum information processing (QIP) and satisfy the DiVincenzo criteria \cite{divincenzo_physical_2000}. State-of-the-art systems recently demonstrated the highest single- \cite{smith_singlequbit_2025} and two-qubit gate fidelities \cite{hughes_trappedion_2025} among all QIP hardware implementations. This excellent control is complemented by a scalable architecture \cite{wineland_experimental_1998, kielpinski_architecture_2002} that increases the number of qubits per processor, and by photonic interconnects to connect multiple quantum processors \cite{monroe_largescale_2014}. Scaling to large numbers of qubits while maintaining such high fidelities and managing the increasing number of required control signals for confinement and shuttling of qubits between registers remains challenging and is a current topic of research \cite{malinowski_how_2023}. Surface-electrode ion traps are well-suited to advance scalability by leveraging the advantages of modern microfabrication methods \cite{chiaverini_surfaceelectrode_2005}. In the commonly used symmetric `five-wire' surface-electrode trap geometry (Figure \ref{fig: inner rect dc}) \cite{chiaverini_surfaceelectrode_2005}, the electrodes producing the electric fields to confine the ions above the surface are all located in the ($x,y$)-plane. Two radio-frequency (RF) electrode stripes provide a radially confining pseudopotential tube along $(x,y=0,z=h_{\rm{ion}})$ where $h_{\rm{ion}}$, the height of the tube above the electrode surface, is set by the width of the RF electrodes and their separation \cite{wesenberg_electrostatics_2008}. 

Additional control electrodes with constant or slowly varying voltages are used to provide confinement along the axial direction $x$ and transport the ions along this direction in the register-based quantum CCD architecture \cite{wineland_experimental_1998,kielpinski_architecture_2002}. Control electrodes can be either placed between (`inner DC electrodes') or outside the RF electrodes stripes (`outer DC electrodes') \cite{nizamani_optimum_2012}. With single-layer microfabrication, only segmented outer control electrodes are feasible. The development of multi-layer fabrication techniques \cite{amini_scalable_2010, maunz_high_2016, bautista-salvador_multilayer_2019, dietl_test_2025} enables increasingly complex multi-layer structures, multi-layer signal routing, and the use of inner control electrodes. Due to their reduced electrode-ion distance, inner control electrodes provide superior control at lower control voltages than outer control electrodes, making them naturally more efficient \cite{blain_hybrid_2021, nizamani_optimum_2012}. While single-layer surface traps and three-dimensional macroscopic traps primarily use rectangular segmented outer control electrodes due to geometrical constraints, this does not imply that this is the ideal shape for inner control electrodes in modern multi-layer surface traps. In previous multi-layer trap designs, rectangular inner control electrodes with axial segmentation (Figure \ref{fig: inner rect dc}), and rectangular inner control electrodes that are additionally split in the radial direction (Figure \ref{fig: inner dc shapes} a)), have been implemented \cite{maunz_high_2016, revelle_phoenix_2020, revelle_roadrunner_2023} and briefly characterized for efficacy (potential curvature per unit voltage) in \cite{blain_hybrid_2021}.

This work extends these findings by exploring non-rectangular shapes for inner control electrodes that provide simultaneous control in both axial and radial directions for ion shuttling and micromotion compensation. The suggested asymmetric inner control electrode shapes eliminate the need for additional outer control electrodes for radial control and micromotion compensation. As a result, the occupied chip area is reduced, increasing the compactness and space efficiency of the entire trap structure while concurrently reducing the number of control signals. The improved control voltage efficiency of using solely inner electrodes enables the device's entire direct-current (DC) supply to be provided by integrated Cryo-CMOS circuits \cite{stuart_chipintegrated_2019, meyer_12_2023, park_cryocmos_2024, meyer_cryogenic_2025}, further enhancing the scalability of surface-electrode trapped-ion quantum processors.

\section{Surface-Electrode Ion Traps}
\label{sec: SE Traps}
For Paul traps, the time-dependent treatment of the radio-frequency field $\vec{E}_{\rm{RF}}\left(\mkern2mu\vec{r},\mkern2mu t \right)= U_{\rm{RF}} \cdot \hat{{E}_{0}}\left( \vecp{r} \right) \cdot \cos{(\Omega_{\rm{RF}} \cdot t)}$ is commonly approximated by the quasi-static pseudopotential \cite{dehmelt_radiofrequency_1968}
\begin{equation}
    \begin{aligned}
      \Phi_{\rm{psd}}\left( \vecp{r} \right) := \frac{q~ U_{\rm{RF}}^2}{4~m~ \Omega_{\rm{RF}}^2 } |\, \hat{E}_{\rm{0}}\left( \vecp{r} \right)|^2~,
    \end{aligned}
    \label{eq: psdeudopotential}
\end{equation}
where $U_{\rm{RF}}$ denotes the RF voltage applied to the RF trap electrodes, $\hat{E}_{0}$ is the normalized electric field per Volt thus produced, $\Omega_{\rm{RF}}$ the RF drive frequency, and $m$ and $q$ the mass and charge of the trapped ion. Combined with the static fields of the $n$ control electrodes $\Phi_{\rm{st}}$, the combined total potential of the trap is given by
\begin{equation}
    \begin{aligned}
      \Phi_{\rm{tot}}\left( \vecp{r} \right) :=&\, \Phi_{\rm{psd}}\left( \vecp{r} \right)
      + \Phi_{\rm{st}}\left( \vecp{r} \right) \\
      =& \frac{q~ U_{\rm{RF}}^2}{4~m~ \Omega_{\rm{RF}}^2 } |\, \hat{E}_{\rm{0}}\left( \vecp{r} \right)|^2 + \sum_{i=1}^{n}  u_{i} ~ \hat{\varphi}_{\rm{st}}^{i}\left( \vecp{r} \right) ~,
    \end{aligned}
    \label{eq: total potential def}
\end{equation}
where $u_{i}$ is the direct-current (DC) voltage applied to the $i$-th control electrode, and $\hat{\varphi}_{\rm{st}}^{i}$ is the normalized potential per Volt thereby produced.

\subsection{Control Electrodes}
\label{sec: control electrodes}
Segmented outer control electrodes are found in many ion surface-electrode traps \cite{seidelin_microfabricated_2006, leibrandt_demonstration_2009, stick_demonstration_2010, moehring_design_2011, ospelkaus_microwave_2011, wright_reliable_2013, mehta_ion_2014, tabakov_assembling_2015, wahnschaffe_singleion_2017, hahn_integrated_2019}. They are inspired by the design of segmented three-dimensional traps and can be realized using single-layer fabrication techniques. Inner DC electrodes require advanced multi-layer fabrication techniques for signal routing \cite{amini_scalable_2010, blain_hybrid_2021}. However, as the electrode-ion distance is shorter for the inner DC electrodes, they can provide finer control and require lower control voltages \cite{blain_hybrid_2021, nizamani_optimum_2012}. Additionally, inner DC electrodes are more space-efficient, thereby enhancing scalability. In the following, we therefore focus on inner control electrodes.

\begin{figure}[h!]
    \centering
    \includesvg[width=0.7\columnwidth]{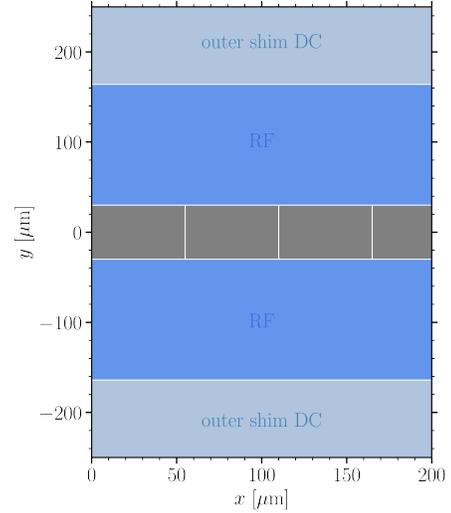}
    \caption{A common five-wire surface electrode trap with symmetric RF electrodes (blue), rectangular inner control electrodes (gray) and additional outer control electrodes (light blue) for radial control.}
    \label{fig: inner rect dc}
\end{figure}

To precisely manipulate a trapped ion, control by static potentials in all three spatial dimensions is necessary to compensate for experimental conditions such as stray fields \cite{berkeland_minimization_1998, allcock_heating_2012, doret_controlling_2012, warring_techniques_2013, hogle_precise_2024}. In typical implementations of inner DC electrodes, these are rectangular and segmented along the axial direction (see Figure \ref{fig: inner rect dc}) \cite{amini_scalable_2010, pino_demonstration_2021, moses_racetrack_2023}. 
This segmentation allows for altering the DC potential along the axial direction, denoted by $x$, thus providing control of the ions in that direction as well as in the out-of-plane radial $z$-direction. As the rectangular inner DC electrodes are not segmented along the in-plane radial $y$-direction and are symmetric with respect to the pseudopotential nodal line ($y=0$), where the ions are trapped, they cannot provide control in the $y$-direction. Therefore, as illustrated in Figure \ref{fig: inner rect dc}, additional outer DC electrodes are required with this configuration.

Another option is to also split the inner DC electrodes along the y-direction (Figure \ref{fig: inner dc shapes} a)) \cite{maunz_high_2016, revelle_phoenix_2020, revelle_roadrunner_2023}, enabling complete radial control without outer DC electrodes. A drawback of this configuration is that it provides a direct line of sight for the ions to the dielectric in the lower layers through the gaps along $y=0$. Usually, such exposure of the ion to dielectrics is avoided, as it can lead to increased heating \cite{harlander_trappedion_2010, bruff_compatibility_2025}.

An alternative proposed in this work is to use asymmetric, non-rectangular electrode shapes to provide simultaneous axial and radial control. To achieve simultaneous control, the inner control electrodes must not be symmetric with respect to the pseudopotential nodal line ($y=0$) and must be axially segmented. Multiple shapes of inner control electrodes meet these requirements. Here, we investigate a selection of elementary shapes shown in Figure \ref{fig: inner dc shapes} b)-f).

\subsection{Gapless Approximation}
\label{sec: gapless}

The electric potentials of surface-electrode traps can be determined using basis functions and the analytic \textit{Gapless Plane Approximation}. The Basis Function technique \cite{hucul_transport_2008} is commonly used for all types of ion traps and was already implicitly introduced in equation \ref{eq: total potential def}. The basis function for the static potentials is the normalized potential per Volt $\hat{\varphi}_{\rm{st}}$. Further, the normalized electric field per Volt $\hat{E}_{0}$ can be expressed as $\hat{E}_{0}= - \nabla \Theta_{\rm{RF}}$, where $\Theta_{\rm{RF}}$ denotes the basis function for the RF (and pseudo-) potential \cite{hucul_transport_2008}. While for three-dimensional traps, typically finite-element methods (FEM) or boundary-element methods (BEM) are used to calculate the basis functions, for surface-electrode traps, the analytic \textit{Gapless Plane Approximation} is applicable \cite{wesenberg_electrostatics_2008, house_analytic_2008, oliveira_biotsavartlike_2001}. The approximation assumes an infinitely large ground plane and neglects the gaps between neighbouring electrodes. In the following, we will use the \textit{Gapless Plane Approximation} implemented in {\tt Python}\textsuperscript{\footnotemark}\footnotetext{\url{https://github.com/nist-ionstorage/electrode}}
to calculate all basis functions and the resulting electric fields and potentials, respectively.

Note that to determine the potential basis function $\hat{\varphi}(\vecp{r})$ for an arbitrarily shaped electrode with area $A_{\rm{DC}}$ in a surface trap, a closed path integral along its oriented boundary $\partial A_{\rm{DC}}$ is calculated \cite{oliveira_biotsavartlike_2001}. Green's Theorem \cite{green_essay_1852,tapp_differential_2016} implies the intuitive relation that the potential is proportional to the electrode's area $\hat{\varphi}(\vecp{r}) \sim A_{\rm{DC}}$.\textsuperscript{\footnotemark}\footnotetext{More precisely, the potential depends on the solid angle $\Omega(\vecp{r})$ spanned by the electrode as seen from the ion's position $\vecp{r}$ \cite{oliveira_biotsavartlike_2001}.} This relation will be essential in the following section, where we ensure that all inner DC electrode shapes have equal areas to make a fair comparison of their potential characteristics.

\section{Characterization of Inner Control Electrode Shapes}
\label{sec: Comparison}
In the following, we characterize the properties of different inner control electrodes. We start by describing the method used to achieve a fair comparison, followed by a characterization of single electrodes held at unit voltage. Finally, we evaluate their performance in realistic use-case scenarios for ion transport and micromotion compensations.

\subsection{Comparing Differently Shaped Control Electrodes}
\label{sec: electrode shapes}
To compare the control electrodes, the specific RF electrode geometry is not crucial. Hence, we use a standard geometry for surface-electrode traps with an ion height of $h_{\rm{ion}} \simeq 70.1~\mu\rm{m}$ and symmetrical RF electrodes, with an RF electrode splitting $A=60~\mu\rm{m}$ and width $B=134~\mu\rm{m}$, similar to \cite{mokhberi_optimised_2017}.

\begin{figure}[h!]
    \centering
    \includesvg[width=1.0\columnwidth]{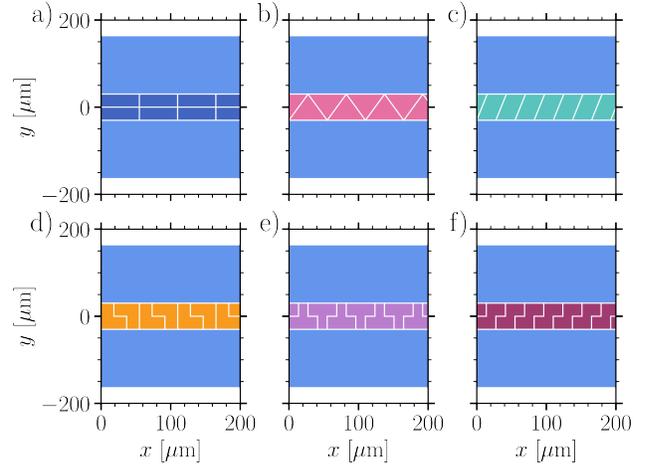}
    \caption{Surface electrode trap layout with a) axial and radial segmented rectangular, b) triangular, c) rhomboid, d) `L'-shaped, e) `T'-shaped, and f) `Z'-shaped inner control electrodes for simultaneous control in axial and radial directions. No additional outer control electrodes are required for radial control. The RF electrodes are shown in blue and the white area outside the RF is ground.}
    \label{fig: inner dc shapes}
\end{figure}

The crucial idea for a fair comparison across different electrode shapes is to normalize the electrode areas, since the potential is proportional to the electrode area $\hat{\varphi}_{\rm{st}} \sim A_{\rm{DC}}$ (see Section \ref{sec: gapless} for details). In previous studies, the optimal width $w_{\rm{ax}}$ for control electrodes has been analyzed for rectangular electrodes \cite{reichle_transport_2006, blain_hybrid_2021}. As the non-rectangular electrode shapes investigated here are not solely defined by the width in the axial direction, this is not a suitable measure. However, we can use it to compute the reference area used for the normalization:

We compute the reference area for a single electrode in the rectangular split configuration shown in Figure \ref{fig: inner dc shapes} a). The optimal axial width $w_{\rm{ax}}$ depends on the effective ion-electrode distance $d$ \cite{reichle_transport_2006, hucul_transport_2008}. For inner control electrodes, the ion-electrode distance is simply the ion height $d=h_{\rm{ion}}=70.1~\mu\rm{m}$. According to \cite{reichle_transport_2006}, we obtain an optimal axial segmentation for $w_{\rm{ax}} = 0.78 \cdot d = 55~\mu\rm{m}$. In the $y$-direction, the width is given by half the RF splitting $A$. We obtain an area of $\hat{A}_{\rm{DC}}=1650~\mu\rm{m}^2$ for a single electrode. We use this as the reference value to normalize the area across all other control electrode shapes for a fair comparison. Note that, therefore, the widths for the non-rectangular shapes differ. The exact widths and parametrization of each of the control electrodes are summarized in Table \ref{tab: parametrization} in the Appendix \ref{sec: appendix parametrization}.

\subsection{Characterization at Unit Voltage}
\label{sec: properties 1V}
In the following, we analyze the fundamental characteristics of the potential $\hat{\varphi}_{\rm{st}}$ (and its derivatives) produced by a single control electrode when held at unit voltage. This characterization allows for a distinct comparison of the different electrode shapes and provides a basis for their effective use in future trap designs. It further promotes the understanding of their collective behavior for ion transport and micromotion compensation in section \ref{sec: transport results}.

\begin{figure}[h!]
    \centering
    \includesvg[width=0.9\columnwidth]{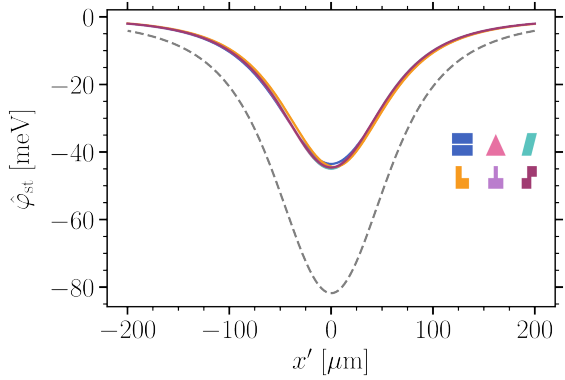}
    \caption{Normalized electric potential $\hat{\varphi}_{\rm{st}}$ for the different inner control electrode shapes held at $-1\,\rm{V}$ relative to the electrodes center located at $x'=0~\rm{ \mu m}$. For reference the potential corresponding to the rectangular inner DC electrode in Figure \ref{fig: inner rect dc} is shown as the gray-dashed line. Note that the area $A_{\rm{DC}}$ of the rectangular electrode is double that of the other electrode shapes, resulting in twice the electric potential ($\hat{\varphi}_{\rm{st}} \sim A_{\rm{DC}}$).}
    \label{fig: 1V dc pot}
\end{figure}

In agreement with previous works, we apply a unit voltage of $-1\,\rm{V}$ to a single control electrode and evaluate the resulting potential relative to the electrode center ($x' = 0$) at a constant ion height $h_{\rm{ion}} =  70.1~\mu\rm{m}$ \cite{reichle_transport_2006,blain_hybrid_2021}. Figure \ref{fig: 1V dc pot} shows the resulting potentials $\hat{\varphi}_{\rm{st}}$ for the different inner control electrode shapes proposed in this study. As intended, the control electrodes produce a potential of the same magnitude, verifying the method introduced in Section \ref{sec: electrode shapes} to normalize the electrode area $A_{\rm{DC}}$. For reference, the potential of the standard rectangular control electrode (see Figure \ref{fig: inner rect dc}) is indicated by the gray-dashed line. Note that the area of the rectangular electrode was \textit{not} normalized, as this would yield a significantly smaller axial width, and therefore generates twice the electric potential.

\begin{figure}[h!]
    \centering
    \includesvg[width=\columnwidth]{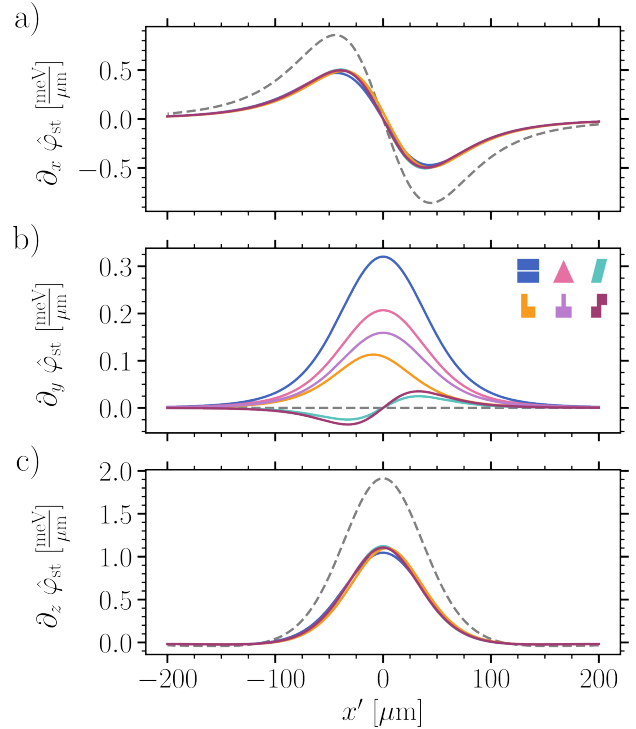}
    \caption{First order derivatives of the normalized electric potential $\partial_i \, \hat{\varphi}_{\rm{st}}$ along the a) $x$-direction, b) $y$-, and c) $z$-direction for the different inner control electrode shapes held at $-1\,\rm{V}$ relative to the electrodes center located at $x'=0~\rm{ \mu m}$. The potentials'  derivatives corresponding to the rectangular inner control electrode in Figure \ref{fig: inner rect dc} are shown as the gray-dashed line for comparison. Due to its symmetry with respect to the $y$-axis the potential derivative $\partial_y \, \hat{\varphi}_{\rm{st}}$ along this direction is zero.}
    \label{fig: 1V dc grad}
\end{figure}

\medskip
Figure \ref{fig: 1V dc grad} shows the corresponding first-order derivatives of the static potential $\partial_i \, \hat{\varphi}_{\rm{st}}$. As the gradient of an electrostatic potential is proportional to the electric field $\vec{E} =-\nabla \varphi$, the first-order derivatives provide insights into the ability to compensate for stray fields. To achieve complete simultaneous axial and radial control, the first-order derivatives must not vanish. It is immediately evident that the standard rectangular inner control electrodes (gray-dashed line) cannot provide this, as the $\partial_y \, \hat{\varphi}_{\rm{st}}$ derivative vanishes. As described in section \ref{sec: control electrodes}, this is due to the lack of asymmetry of the rectangular electrode shape in the $y$-direction. As expected, all other proposed asymmetric electrode shapes achieve such simultaneous control. The magnitude of the potential derivative indicates how demanding the control in the specific direction is. In the axial ($x\text{-)direction}$ and radial out-of-plane ($z$-)direction, the asymmetric electrodes perform equally well. For all of them, providing control in the $z$-direction is the least demanding, followed by the axial control. This capability derives from the electrode layout, in which all inner electrodes are placed directly underneath the trapping position ($y=0$, $z=h_{\rm{ion}}$) and have similar axial segmentation $w_{\rm{ax}}$. The control along the $y$-direction, parallel to the surface, is the most demanding and varies significantly among the asymmetric electrode shapes. The best control in this direction is achieved with the axially and radially segmented rectangular electrode (blue). The large $\partial_y \, \hat{\varphi}_{\rm{st}}$ derivative arises from the significant geometric asymmetry introduced by its explicit segmentation along the $y$-axis. The triangular, `T'- and `L'-shaped electrodes provide adequate control as well. Notably inferior control is given by the point-symmetric rhomboid and `Z'-shaped electrodes.

\begin{figure}[h!]
    \centering
    \includesvg[width=\columnwidth]{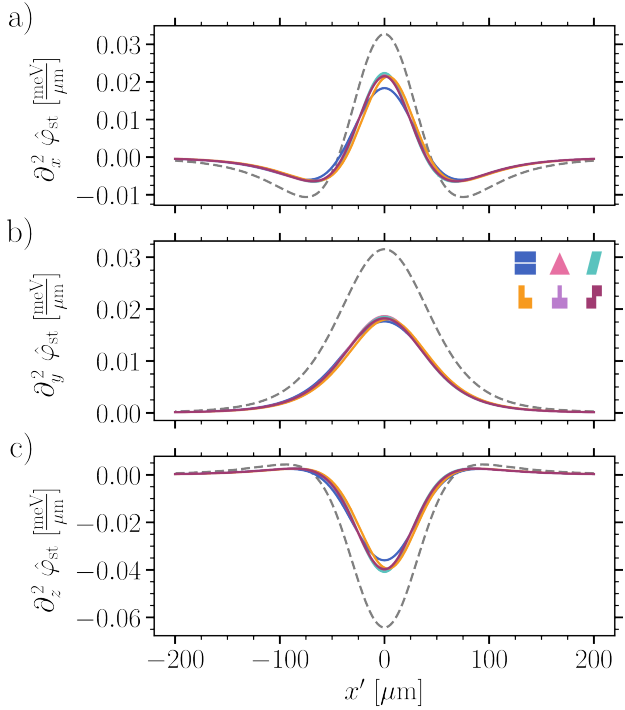}
    \caption{Second order derivatives of the normalized electric potential $\partial_i^{\,2} \, \hat{\varphi}_{\rm{st}}$ along the principal axis $\rm{tr}(H(\hat{\varphi}_{\rm{st}}))$ for the different inner control electrode shapes held at $-1\,\rm{V}$ relative to the electrodes center located at $x'=0~\rm{ \mu m}$. The potential's derivatives corresponding to the rectangular inner control electrode in Figure \ref{fig: inner rect dc} are shown as gray-dashed lines for reference.}
    \label{fig: 1V-dc-hessian-trace}
\end{figure}

\begin{figure}[h!]
    \centering
    \includesvg[width=\columnwidth]{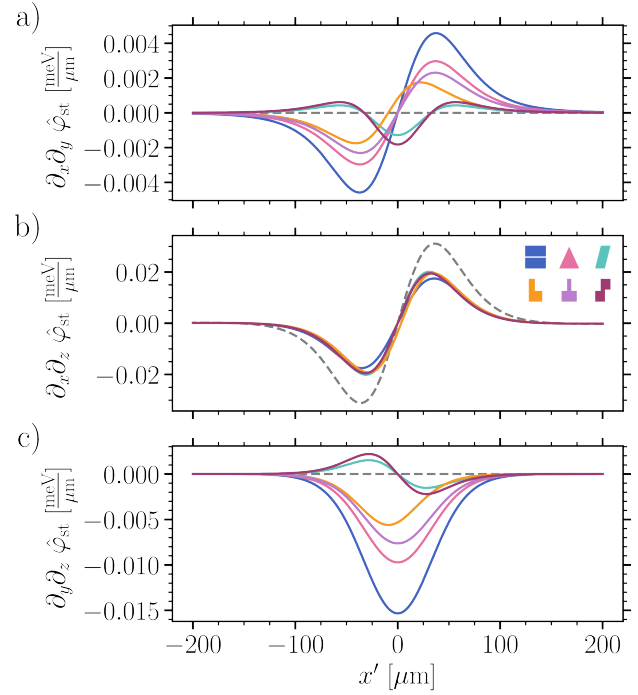}
    \caption{Mixed second order derivatives of the normalized electric potential $\partial_i\partial_j \, \hat{\varphi}_{\rm{st}}$ for the different inner dc electrode shapes held at $-1\,\rm{V}$ relative to the electrodes center located at $x'=0~\rm{ \mu m}$. The potential's derivatives corresponding to the rectangular inner DC electrode in Figure \ref{fig: inner rect dc} are shown as gray-dashed lines for reference. Due to symmetry with respect to the $y$-axis, the potential derivatives $\partial_x\partial_y \, \hat{\varphi}_{\rm{st}}$ and $\partial_y\partial_z \, \hat{\varphi}_{\rm{st}}$ vanish.}
    \label{fig: 1V-dc-hessian-mixed}
\end{figure}

\medskip
The second-order derivatives of the potential $\partial_i\partial_j \, \hat{\varphi}_{\rm{st}}$ yield the curvatures at the trapping position, which define the corresponding motional mode frequencies and oscillation directions of the ion.
Second-order derivatives are collectively represented by the Hessian matrix $H_{\rm{st}}$. In Figure \ref{fig: 1V-dc-hessian-trace}, the derivatives of the Hessian main diagonal $tr(H_{\rm{st}})$ are shown, which represent the motional mode frequencies $\omega_{\rm{st},\, i}$ along the principal axes via $\omega_{\rm{st},\, i}^2 = \frac{q}{m} \, \partial_i^{\,2} \, \hat{\varphi}_{\rm{st}}$. Typically, in a linear Paul trap, the mode frequencies along the radial directions ($y$ and $z$) are provided primarily by the RF potential. Therefore, the most relevant capability for the inner DC electrodes is to generate curvature along the axial ($x$-)direction shown in Figure \ref{fig: 1V-dc-hessian-trace}~a) corresponding to the axial frequency $\omega_{\rm{st},\, x}$. Here, the different control electrode shapes perform similarly due to their similar axial widths $w_{\rm{ax}}$. Yet the axial and radial segmented rectangular electrodes (blue) produce notably less axial curvature.

The radial curvatures generated by the static potential can be understood as the ability to modify those generated by the RF potential. Due to Laplace's equation $\Delta\, \hat{\varphi}_{\rm{st}} =0$, the $\partial_z^{\,2}$ curvatures are of opposite sign and have about twice the magnitude compared to the $\partial_x^{\,2}$ and $\partial_y^{\,2}$ curvatures. This behavior, observed across all electrode shapes, results from the planar electrode geometry of surface-electrode ion traps.

\medskip
The off-diagonal terms of the Hessian can be used to generate rotations of the oscillation basis, for example, necessary for laser cooling or manipulating ion crystals. Since the derivatives commute $\partial_i\partial_j \, \hat{\varphi}_{\rm{st}} = \partial_j\partial_i \, \hat{\varphi}_{\rm{st}}$, Figure \ref{fig: 1V-dc-hessian-mixed} displays only the mixed derivatives corresponding to the upper triangular of the Hessian. When they are non-zero, rotations can be created with the static potential. The simple rectangular inner control electrodes (gray-dashed line) can not produce $x$-$y$ and $y$-$z$ rotations due to the lack of asymmetry in the $y$-direction. All proposed asymmetric inner control electrodes can provide the complete set of rotations. However, there are differences in the magnitude between the various electrode shapes and cross-derivatives. Overall, the $\partial_x\partial_y$ and $\partial_y\partial_z$ derivatives are an order of magnitude smaller than the $\partial_x\partial_z$ derivatives, indicating they are more costly to create. This observation is consistent with the increased magnitude of the $\partial_z^{\,2}$ derivative, suggesting predominant contributions from the $z$-direction originating from the planar geometry of surface traps. All asymmetric electrode shapes produce a similar $\partial_x\partial_z$ derivative but vary in the attainable $\partial_x\partial_y$ and $\partial_y\partial_z$ derivatives. Similar to the first-order derivatives, the axially and radially segmented rectangular electrodes (blue) provide the highest derivatives, followed by the triangular, `T'-, and `L'-shaped electrodes, and with significantly smaller derivatives by the point-symmetric rhomboid and `Z'-shaped electrodes.

Overall, the analysis of the potential characteristics at unit voltage suggests that the axially and radially segmented rectangular electrodes (blue) are best suited, and the triangular, `T'-, and `L'-shaped electrodes should provide decent results for ion transport and micromotion compensation. Nevertheless, we evaluate the performance of all the electrode shapes in the following sections.

\subsection{Performance for Ion Transport and Micromotion Compensation}
\label{sec: transport results}

\begin{table*}[h]
    \centering
    \begin{tabular}{lcccccccc}
        \toprule
        \raisebox{0.4\height}{\textbf{Electrode Shape}} & \includesvg[width=0.5cm,height=0.5cm,keepaspectratio]{plain_rectangular.svg} & \includesvg[width=0.5cm,height=0.5cm,keepaspectratio]{plain_rectangular_split.svg} & \includesvg[width=0.5cm,height=0.5cm,keepaspectratio]{plain_triangular.svg} & \includesvg[width=0.5cm,height=0.5cm,keepaspectratio]{plain_L.svg} & \includesvg[width=0.5cm,height=0.5cm,keepaspectratio]{plain_T.svg} & \includesvg[width=0.5cm,height=0.5cm,keepaspectratio]{plain_Z.svg} & \includesvg[width=0.5cm,height=0.5cm,keepaspectratio]{plain_rhomboid.svg} \\
        \midrule
        \multicolumn{9}{c}{\textbf{Ion Transport}} \\
        \midrule
        $\max. \,|V|$ & 0.126 & 0.126 & 0.123 & 0.151 & 0.128 & 0.357 & 0.447 & $[\rm{V}]$ \\
        $ \Delta \Phi_{\rm{tot},\,\rm{ax}\,}\big|_{x\,=\,x_{0} \pm 150\,\mu \rm{m}}$ & 3.2 & 3.2 & 4.4 & 3.3 & 4.2 & 19 & 21 & $[\rm{meV}]$ \\
        \midrule
        \multicolumn{9}{c}{\textbf{Micromotion Compensation}} \\
        \midrule
        $\mathbf{x:}~$ $\max. \,|V|$ & 0.188 & 0.214 & 0.185 & 0.228 & 0.186 & 10 & - & $[\rm{V}]$ \\
        $\mathbf{y:}~$ $\max. \,|V|$ & - & 0.801 & 1.24 & 2.75 & 1.6 & - & - & $[\rm{V}]$ \\
        $\mathbf{z:}~$ $\max. \,|V|$ & 0.0402 & 1.07 & 0.848 & 0.942 & 0.841 & 1.51 & 2.49 & $[\rm{V}]$ \\
        \bottomrule
    \end{tabular}
    \caption{Key results for ion transport and micromotion compensation for the different inner control electrode shapes. For all proposed asymmetric inner control electrodes, a satisfying solution for the transport voltages was found. The magnitudes of the required control voltages $\max. \,|V|$ indicates how easily transport or shim compensation can be realized. The transport potentials are further assessed by their axial trap depth $ \Delta \Phi_{\rm{tot},\,\rm{ax}\,}\big|_{x\,=\,x_{0} \pm 150\,\mu \rm{m}}$. Not all micromotion compensation fields (`shim fields') could be obtained by some of the inner control electrodes indicated by `-'. The standard rectangular electrodes (gray) can not provide shim fields in the $y$-direction due to their symmetric shape. The rhomboid and `Z'-shaped electrodes provide too little control for specific shim fields. These results are consistent with the previous characterization at unit voltage. For details, see the text.}
    \label{tab: results}
\end{table*}

In addition to the characterization at unit voltage, we evaluate the performance of the different inner control electrodes under realistic conditions for ion transport and micromotion compensations.

\subsubsection{Ion Transport}
As a test environment, we use a $10~\rm{mm}$ long linear surface trap with the same RF geometry as in the previous section to approximate an ideal, infinitely long linear trap with minimal pseudopotential variations. The control electrodes cover the innermost $\pm 1~\rm{mm}$ of the trap in the configuration shown in Figure \ref{fig: inner dc shapes}. Ion transport is calculated from $-250~\mu\rm{m}$ to $+250~\mu\rm{m}$ in $5~\mu\rm{m}$ steps, to avoid finite-length effects of the DC structure, and covers multiple control electrode widths $w_{\rm{ax}}$. All calculations are performed for \BeIon ions, an RF voltage  $U_{\rm{RF}}=75~\rm{V}$ (zero-peak), and an RF frequency $\Omega_{\rm{RF}}=2\pi \cdot 88.191~\rm{MHz}$. Nonetheless, the results would be similar for another selected ion species. A typical axial frequency for \BeIon is $\omega_{x}=2\pi \cdot 1~\rm{MHz}$, which we use for the ion transport calculations. Moreover, standard constraints with vanishing electric-field components $\partial_{i}\,\Phi_{\rm{tot}}=0$ and vanishing off-diagonal Hessian components $\partial_{i}\partial_{j}\,\Phi_{\rm{tot}}=0$ where $i,j \in \{x,y,z\}$ are used.\textsuperscript{\footnotemark}\footnotetext{For the rectangular electrodes (gray), we need to drop the $\partial_{y}$, $\partial_{x}\partial_{y}$, and $\partial_{y}\partial_{z}$ constraints, as they do not provide control over these derivatives due to their symmetric shape (cf. section \ref{sec: properties 1V}).}
Additionally, we limit the calculations to the 12 control electrodes that generate the highest potential at the current ion position for a unit voltage of $1~\rm{V}$ at each step of the transport. The other \textit{inactive} electrodes are set to $0~\rm{V}$. We thereby localize the required electrodes to the relevant chip area, which is particularly useful for larger ion-trap systems and enables parallel shuttling operations.\textsuperscript{\footnotemark}\footnotetext{The choice of 12 \textit{active} electrodes is somewhat arbitrary but has been found to work well. Mathematically, since the voltages applied to the control electrodes represent the degrees of freedom $N$, the number of constraints $C$ must be at least as large as the number of degrees of freedom $N \geq C$ to obtain a solution to the linear system of equations.}
To ensure experimentally realizable voltages, we set a voltage limit of $\pm 10~\rm{V}$ for the \textit{active} electrodes, which is a typical range for Digital-to-Analog Converters (DACs) used in trapped-ion experiments.\textsuperscript{\footnotemark}\footnotetext{\url{https://sinara.technosystem.pl/modules/fastino/}}

To compute the transport voltages, we use the standard solver of the {\sc electrode}\textsuperscript{\footnotemark}\footnotetext{\url{https://github.com/nist-ionstorage/electrode}} package, which uses the {\sc cvxopt}\textsuperscript{\footnotemark}\footnotetext{\url{https://cvxopt.org/index.html}} library to solve convex matrix problems via linear programming. 
It effectively minimizes a weighted-sum objective based on the $1$-norm of the voltages $V$ and their derivatives subject to the specified constraints:

\begin{mini}[l]
    { } %variable
    {w_0 \cdot {\left\lVert V \right\rVert}_1 + w_2 \cdot{\left\lVert \partial^2\,V \right\rVert}_1} %objective function
    {}
    {}
\end{mini}
We use a weight $w_0=1\cdot10^{-6}$ for the $0$\textsuperscript{th}-order voltages and, in addition, a weight $w_2=2\cdot10^{-6}$ for the $2$\textsuperscript{nd}-order voltage derivatives to improve the smoothness of the voltages. Both weights were found to work satisfactorily by manual testing. The solver tolerance was set to $1 \cdot10^{-6}$.

\medskip
The key transport results are summarized in Table \ref{tab: results}. For all proposed asymmetric inner control electrodes, a satisfying solution for the transport voltages was found. For comparison, the results for standard rectangular electrodes (gray), obtained with relaxed constraints, are shown. To evaluate the result, a straightforward metric is the magnitude of the required control voltages $\max. \,|V|$. Since the solver minimizes the control voltages $V$, their magnitude indicates how demanding it is to meet the transport constraints. As indicated by the characterization at unit voltage in the previous section \ref{sec: properties 1V}, the axially and radially segmented rectangular (blue), triangular, `T'-, and `L'-shaped control electrodes perform similarly well with comparable maximum transport voltages. In comparison, the rhomboid and `Z'-shaped electrodes require about three times higher voltages. Such increased control voltages are consistent with their characteristics at unit voltage - due to their point-symmetric shape, they have difficulty collectively canceling the second-order mixed derivatives $\partial_{i}\partial_{j}$ included in the transport constraints.

Further, we can assess the transport potentials by the resulting axial trap depth $ \Delta \Phi_{\rm{tot},\,\rm{ax}\,}\big|_{x\,=\,x_{0} \pm 150\,\mu \rm{m}}$. To determine the axial trap depth, we calculate the potential in an interval around the trapping position $x_0$ along the axial ($x\text{-)direction}$. When the interval is too small, the axial trap depth is underestimated. We obtain reliable results with a range of $150~\mu\rm{m}$. Note that the axial trap depths obtained here are the bare results of the simple transport constraints and by no means optimized. Additional efforts can be employed to improve the axial trap depths. Yet they portray the general behavior that we analyze here. The standard rectangular (gray), radially segmented rectangular (blue), and `L'-shaped control electrodes generate similar axial trap depths. For the triangular and `T'-shaped control electrodes, the depth is increased by more than $30\%$. The rhomboid and `Z'-shaped electrodes produce substantially larger axial trap depths, due to the increased control voltages.

\subsubsection{Micromotion Compensation}
In addition to ion transport, we evaluate the inner control electrodes' ability to generate micromotion compensation (`shim') fields in all spatial directions. The shim fields are calculated covering the entire control electrode width $w_{ax}$ in $1~\mu\rm{m}$ steps along the axial direction. Therefore, the results are not position-dependent, whereas selecting a single position would introduce bias. We use the same solver, $12$ active electrodes, a $\pm 10\rm{V}$ voltage limit, and the same weights for the voltage derivatives as before. However, the constraints for the shims are different. Since the shim fields are static fields, we omit the pseudopotential. We normalize the shim field strength to $100~\rm{V/m}$. Thereby, the shim fields serve as basis functions that can be scaled in the experiment. We constrain the static potential to generate this field at the ion position in just the desired direction $\partial_i \,\Phi_{\rm{st}}= 100~\rm{V/m}$ and $\partial_{j \neq i} \,\Phi_{\rm{st}}= 0~$. Further, to obtain \textit{pure} shim fields that do not induce additional curvatures, we constrain all second-order derivatives $\partial_{i}\partial_{j}\,\Phi_{\rm{st}}=0$ except the $\partial_z^{\,2}$-derivative, which implicitly vanishes due to Laplace's equation.

As the rectangular electrodes (gray) cannot generate derivatives in the $y$-direction due to their symmetric shape, we do not calculate the $y$ shim field, indicated by the dash `-' in Table \ref{tab: results}. For the same reason, we also have to drop the $\partial_{y}$, $\partial_{x}\partial_{y}$, and $\partial_{y}\partial_{z}$ constraints when calculating the $x$- and $z$-shims for this electrode shape. Relaxing these constraints results in deviations of up to $5~\rm{kHz}$ in the radial mode frequencies. For the other asymmetric electrode shapes, such deviations were not observed due to the complete set of constraints. The radially segmented rectangular (blue), triangular, and `T'-shaped control electrodes provide all necessary shim fields with similar voltage magnitudes. Slightly larger voltages are required with the `L'-shaped electrodes, in particular for the $y$ shim field. The rhomboid and `Z'-shaped electrodes offer too little control for some shim fields, indicated by the dash `-' in Table \ref{tab: results}. \newline The `Z'-shaped electrodes max out the $\pm 10\rm{V}$ limit for the $x$-shim and could only generate $y$-shim without this limit and extreme voltages ($>500~\rm{V}$). The rhomboid electrodes could not create any $x$- and $y$-shim even without the voltage limit. These results are consistent with the observation for the characterization at unit voltage, where the rhomboid and `Z'-shaped electrodes generated the smallest $y$-derivatives (cf. Figure \ref{fig: 1V dc grad}) and exhibited significantly smaller $\partial_{x}\partial_{y}$ and $\partial_{y}\partial_{z}$ derivatives (cf. Figure \ref{fig: 1V-dc-hessian-mixed}). Due to their point-symmetric shape, the electrodes have equal areas above and below the trap axis ($y=0$). As seen in the transport, they therefore have difficulty collectively canceling the second-order mixed derivatives and can not generate \textit{pure} $x$- and $y$-shim fields. Yet they can still provide $z$-shim fields, albeit at higher voltages than the other shapes.

\section{Conclusion and Outlook}
\label{sec: Discussion}
We have discussed novel asymmetric inner control electrodes for simultaneous axial and radial control in multi-layer surface-electrode ion traps, eliminating the need for commonly used additional outer control electrodes. We developed and verified a method to compare the static potentials generated by prototypical control electrode shapes by normalizing their areas. We have characterized the static potential and its derivatives up to second order at unit voltage, extending previous work on rectangular control electrodes by \cite{reichle_transport_2006} and \cite{blain_hybrid_2021}. We have further tested the proposed asymmetric inner control electrodes in two realistic use-case scenarios for ion transport and micromotion compensation in a linear surface-trap model. Consistent results from unit-voltage characterization, ion transport, and micromotion compensation confirm that asymmetric inner control electrodes can indeed provide simultaneous control in all directions. Overall, the best results were obtained with the established radially segmented rectangular electrodes (blue) and the newly introduced triangular electrodes. Decent results were also obtained with the new `T'- and `L'-shaped electrodes. The rhomboid and `Z'-shaped electrodes could provide satisfying linear transport with increased voltages. However, they could not produce all the necessary micromotion compensation fields due to their point-symmetric shape. They might still be used with additional outer control electrodes or relaxed constraints on the compensation fields.

In conclusion, radially segmented rectangular (blue), triangular, `T'- and `L'-shaped inner control electrodes appear to be most promising in surface-electrode ion traps. Our fundamental characterization of these electrodes lays the groundwork for future chip designs and experimental applications. By eliminating the need for commonly used additional outer control electrodes, asymmetric inner control electrodes increase the compactness and space efficiency of surface-electrode traps and, concurrently, reduce the number of control signals. Using solely inner electrodes further improves the control-voltage efficiency and enables the device's entire direct-current (DC) supply to be provided by integrated Cryo-CMOS circuits \cite{stuart_chipintegrated_2019, meyer_12_2023, park_cryocmos_2024, meyer_cryogenic_2025}. All-together asymmetric inner control electrodes enhance the scalability of surface-electrode trapped-ion quantum processors based on the QCCD architecture. Similarly, surface-electrode Penning traps \cite{jain_penning_2024}, stacked-wafer traps \cite{decaroli_design_2021, auchter_industrially_2022}, 3D-printed micro traps \cite{xu_3dprinted_2025}, or arrays for quantum simulation \cite{holz_2d_2020, valentini_demonstration_2025} could benefit from adapted control-electrode implementations.

\medskip
In this first characterization, we selected a subset of asymmetric electrode shapes, which is certainly not complete. Future work can explore other asymmetric electrode shapes or optimize the geometric parameters of the presented shapes. Further, it would be possible to formulate this as an independent shape optimization problem. Similarly, control electrodes could be optimized for other operations, such as splitting \& merging of ion crystals, rotations of single ions or ion crystals, and reordering (position swapping) of ions, or combinations of these. For specific use cases, a characterization of higher-order derivatives of the static potentials might also be of interest. Additional work could explore control electrode shapes in traps with integrated photonics \cite{mehta_integrated_2016} or embedded conductors for microwave-driven quantum-logic gates \cite{ospelkaus_trappedion_2008,ospelkaus_microwave_2011}.

\section*{Acknowledgments}
The authors thank Rodrigo Munoz and Giorgio Zarantonello for helpful discussions.

\section*{Funding}
We acknowledge funding by the Quantum Valley Lower Saxony (QVLS) Q-1 project, the German Federal Ministry of Research, Technology and Space (BMFTR) through the MIQRO and ATIQ projects, and the European Union via the Millenion-SGA1 project (HORIZON-CL4-2022-QUANTUM-01-SGA).

\printbibliography

\clearpage
\appendix
\section*{Appendix}
\addcontentsline{toc}{section}{Appendix}
\renewcommand{\thesubsection}{\Alph{subsection}}
\subsection{Parametrization of the Inner Control Electrodes}
\label{sec: appendix parametrization}

Table \ref{tab: parametrization} illustrates the parametrization of the different inner control electrodes. The electrode shapes are commonly described by their axial width $w_{\rm{ax}}$ and, in some cases, by an additional geometric parameter ($\alpha$ or $\varepsilon$). In the $y$-direction, the width of the inner control electrodes is constrained by the splitting of the RF electrodes $A = 60~\mu\rm{m}$, which is fixed by the RF electrode geometry. As discussed in Section \ref{sec: Comparison}, the area of the inner control electrodes is normalized to $\hat{A}_{\rm{DC}}=1650~\mu\rm{m}^2$ to ensure proper comparison. Therefore, the width in the axial direction $w_{\rm{ax}}$ was adapted for each electrode accordingly. The exact width of each electrode is given in Table \ref{tab: parametrization}.
Additionally, to parametrize the rhomboid electrode shape, the angle $\alpha$ is introduced. It effectively controls the degree of asymmetry in the $y$-direction. Similarly, the parameter $\varepsilon$ determines the exact geometry of the `L', `T', and `Z'-shaped electrodes. Since this article addresses the general use of asymmetric inner control electrodes and their characteristics for a selection of differently shaped control electrode types, the geometric parameters $\alpha$ and $\varepsilon$ were arbitrarily chosen. Regardless, optimizing the geometric parameters for a particular control electrode type is possible but would exceed the scope of this article.

\begin{table}[h!]
\centering
\begin{tabular}{cc>{\centering\arraybackslash}m{2.5cm}}
\toprule
Axial Width & Geom. Parameter & Shape \\ 
\midrule

$w_{x}=55~\mu\rm{m}$ & - &
\includesvg[width=2.5cm]{parametrization_rectangular_split.svg} \\

$w_{x}=55~\mu\rm{m}$ & - &
\includesvg[width=2.5cm]{parametrization_triangular.svg} \\

$w_{x}=27.5~\mu\rm{m}$ & $\alpha = 30^{\circ}$ &
\includesvg[width=2.5cm]{parametrization_rhomboid.svg} \\

$w_{x}=36.7~\mu\rm{m}$ & $\varepsilon = \frac{1}{2}$ &
\includesvg[width=2.5cm]{parametrization_L.svg} \\

$w_{x}=41.2~\mu\rm{m}$ & $\varepsilon = \frac{1}{3}$ &
\includesvg[width=2.5cm]{parametrization_T.svg} \\

$w_{x}=27.5~\mu\rm{m}$ & $\varepsilon = \frac{1}{2}$ &
\includesvg[width=2.5cm]{parametrization_Z.svg} \\

\bottomrule
\end{tabular}
    \caption{Parametrization of the inner control electrode shapes with normalized area $\hat{A}_{\rm{DC}}=1595~\mu\rm{m}^2$. The axial width $w_{\rm{ax}}$ was optimized for the rectangular shape according to \cite{reichle_transport_2006}. The other widths differ as we normalized the area for a fair comparison. The width in the $y$-direction is given by the RF electrode splitting $A$. The geometric parameters $\alpha$ and $\varepsilon$ were chosen arbitrarily.}
    \label{tab: parametrization}
\end{table}

\end{document}